# TimeTrader: Exploiting Latency Tail to Save Datacenter Energy for On-line Data-Intensive Applications

Balajee Vamanan  
Purdue University  
bvamanan@ecn.purdue.edu

Hamza Bin Sohail  
Purdue University  
hsohail@ecn.purdue.edu

Jahangir Hasan  
Google Inc.  
jahangir@google.com

T. N. Vijaykumar  
Purdue University  
vijay@ecn.purdue.edu

**Abstract**

*Datacenters running on-line, data-intensive applications (OLDIs) consume significant amounts of energy. However, reducing their energy is challenging due to their tight response time requirements. A key aspect of OLDIs is that each user query goes to all or many of the nodes in the cluster, so that the overall time budget is dictated by the tail of the replies' latency distribution; replies see latency variations both in the network and compute. Previous work proposes to achieve load-proportional energy by slowing down the computation at lower datacenter loads based directly on response times (i.e., at lower loads, the proposal exploits the average slack in the time budget provisioned for the peak load). In contrast, we propose TimeTrader to reduce energy by exploiting the latency slack in the sub-critical replies which arrive before the deadline (e.g., 80% of replies are 3-4x faster than the tail). This slack is present at all loads and subsumes the previous work's load-related slack. While the previous work shifts the leaves' response time distribution to consume the slack at lower loads, TimeTrader reshapes the distribution at all loads by slowing down individual sub-critical nodes without increasing missed deadlines. TimeTrader exploits slack in both the network and compute budgets. Further, TimeTrader leverages Earliest Deadline First scheduling to largely decouple critical requests from the queuing delays of sub-critical requests which can then be slowed down without hurting critical requests. A combination of real-system measurements and at-scale simulations shows that without adding to missed deadlines, TimeTrader saves 15-19% and 41-49% energy at 90% and 30% loading, respectively, in a datacenter with 512 nodes, whereas previous work saves 0% and 31-37%. Further, as a proof-of-concept, we build a rack-scale real implementation to evaluate TimeTrader and show 10-30% energy savings.*

## 1  Introduction

Datacenters host many of modern Internet services today such as Web Search, social networking, e-commerce, and cloud computing. Datacenters consume tens of megawatts of electric power [8], which accounts for millions of dollars in annual operating costs [30]. Of their total power, modern datacenters spend about 10% on cooling and power distribution overheads (their Power Usage Effectiveness is 1.12 [15]) and about 5% on networking equipment, leaving about 85% for servers of which memory and disk take up 45% and processors consume 55% (i.e., 47% of total) [8, 15, 23]. TimeTrader focuses on the substantial processor power.

Many of Internet services are provided by on-line, data-intensive applications (OLDIs) which often process vast amounts of Internet data (e.g., Web Search and Key-Value stores) [25]. Such services typically operate under tight response time budgets set by service-level agreements (SLAs) (e.g., 200 ms for a Web Search query) [16]. Processing of a query often involves hundreds or thousands of servers working in parallel on memory-resident data [7, 11]. OLDIs have two distinguishing characteristics. (1) They employ a multi-level tree-like software architecture where each query goes to *all or many* leaves. Consequently, though only a few leaves' replies are slow, the overall SLA budget is dictated by the tail of the leaves' reply latency distribution [11] (e.g., the 99.9$^{th}$ percentile leaf latency in a 1000-leaf tree). Replies arriving after the deadline are dropped for responsiveness. (2) Both the network and compute at the leaf contribute to significant variability in the latency of the leaves' replies, as we explain in Section 2.1 (e.g., a request or reply takes 2-30 ms in the network [5, 37, 38] and leaf computation takes 40-120 ms [34]). Both network and compute variations occur at all datacenter loads though the spread is greater at higher loads.

Using low-power or sleep modes is a common approach to saving energy. Unfortunately, OLDIs' time budgets and inter-arrival times are too short for the transition latencies of low-power modes [24, 25]. As such, the low-power modes would incur many deadline violations [23]. Alternately, an insightful recent work, called Pegasus [23], achieves load-proportional energy by slowing down the leaf computation at lower datacenter loads while carefully ensuring that SLAs are not violated (e.g., at night times [25]). Pegasus exploits the mean slack at lower loads in the time budget provisioned for the peak load.

In contrast, we propose *TimeTrader* to reduce energy by exploiting sub-critical leaves' latency slack (e.g., 80% of leaves in *every* query complete within a 3$^{rd}$-4$^{th}$ of the budget.). This slack is present at all loads (modern datacenters operate at high loads during the day [25]); and subsumes Pegasus' load-related slack. Pegasus exploits the mean load-related slack, common to all leaves at lower loads, to *shift* the response time distribution. Instead, TimeTrader *reshapes* the response time distribution at all loads by slowing down individual sub-critical leaves so that they are closer to, but within, the deadline than the default distribution. While TimeTrader saves more energy than Pegasus at low loads, TimeTrader achieves significant savings even at the peak load, which occurs often and where Pegasus has no opportunity. Thus, TimeTrader converts the performance disadvantage of latency tails [11] into an energy advantage.

TimeTrader employs two ideas. First, TimeTrader trades time across system layers, borrowing from the network layer and lending to the compute layer. Each query results in a request-compute-reply-aggregate sequence where the requests from parents to the leaves and replies from the leaves to their parents see variability in the network, and the compute phase sees variability in the leaf server. OLDIs break up the total time budget into a component each for request, compute, reply, and aggregate. We make the *key* observation that because request comes before compute, the slack in faster requests can be transferred to their corresponding compute without any prediction or risk of missing the deadline. To exploit the variations in compute, we make the key observation that while Pegasus captures average variations due to datacenter-wide load changes, each individual query's queuing at the leaf server varies significantly even under a fixed load providing more opportunity (e.g., due to "instantaneous" variations in work and load). Unlike request and compute-queuing, unfortunately, reply comes after compute and reply latency is unpredictable due to the highly-timing-dependent nature of network latencies (Section 2.1). Therefore, the slack in faster replies cannot be transferred easily to their compute. As such, TimeTrader exploits the request and compute slacks but not the reply slack.

Second, despite the slack, such slowing down is challenging in the presence of long tails and SLA guarantees. Even though a sub-critical request has slack, slowing it down may hurt another, critical request that is queued behind the sub-critical request. To address this issue, we leverage the well-known idea of Earliest Deadline First (EDF) scheduling [22] to decouple critical requests from the queuing delays of sub-critical requests by placing the former ahead of the latter in the leaf servers' queues. Conventional implementations and Pegasus cannot exploit EDF because they do not distinguish between critical and sub-critical requests. Due to its decoupling, EDF pulls in the tail and reshapes the leaves' response time distribution (without improving the mean), enabling TimeTrader to use the *per-leaf* slack to shift further the distribution closer to the deadline than with network slack alone. Though this shift lengthens the mean service time, such an increase does not worsen throughput. Because OLDIs' response times are sensitive to tail latencies, compute-queuing delays are kept low even at high loads via high throughput-parallelism (i.e., there is compute-throughput slack even at high loads). As such, TimeTrader's longer service times tap into this throughput slack without causing loss of throughput.

Finally, TimeTrader employs two key mechanisms to realize the above ideas. Transferring the request slack from the network to the compute is challenging due to lack of fine-grained (sub-ms) synchronization between a parent and the leaves. To address this issue, we leverage the well-known Explicit Congestion Notification (ECN) in IP [32] and TCP timeouts to inform the leaves whether a request encountered timeout or congestion in the network and hence does not have slack. Further, because the slack lengths are tens of milliseconds, we use power management schemes with response times of 1 ms, similar to Pegasus (e.g., Running Average Power Limit (RAPL) [1]).

In summary, the paper's contributions are:

- TimeTrader reshapes the response time distribution at all loads by slowing down individual sub-critical leaves without increasing SLA violations;
- TimeTrader exploits the request and compute slack on a per-leaf, per-query basis;
- TimeTrader leverages EDF to largely decouple critical requests from the slowing down of sub-critical requests; and
- TimeTrader leverages (a) network signals such as TCP timeouts and ECN to circumvent the lack of fine-grained synchronization between parent and leaves and (b) modern, low-latency power management to fit within OLDI timescales.

Using a combination of real-system measurements and at-scale simulations, we show that without adding to missed deadlines TimeTrader saves 15-19% and 41-49% energy at 90% and 30% loading, respectively, in a datacenter with 512 nodes, whereas previous work saves 0% and 31-37%. We also build a rack-scale real implementation to evaluate TimeTrader and show 10-30% energy savings.

The rest of the paper is organized as follows. Section 2 describes the background and the challenges. Section 3 describes TimeTrader's details. Section 4 describes our experimental methodology and Section 5 and 6 present our results. Section 7 discusses related work. Finally, Section 8 concludes the paper.

## 2 Challenges and opportunities

### 2.1 Background

As discussed, OLDIs typically employ a tree-based software architecture where the data to be queried resides in the leaf nodes' memory for fast access [7, 11] (see Figure 1). For instance, in Web Search and Key-Value store, the search index and the key-value pairs are partitioned across the leaves in a well load-balanced manner (e.g., using good hashing). In Web Search, every query is broadcast to all the leaves whose results are aggregated based on some ranking scheme (e.g., Google's PageRank). Typical use of key-value stores involve looking up several keys, so that each top-level request generates lookups in several hundreds of leaves, as noted in [23] (e.g., a user's Facebook page typically comprises of several hundreds of objects).

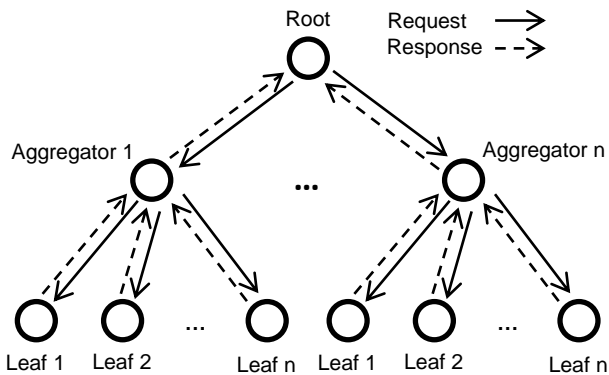

**Figure 1: OLDI software architecture**

Each query involves a request-compute-reply-aggregate sequence where the query generates requests to the leaves going through multiple levels in the tree (see Figure 1); each leaf looks up its memory to compute its result and sends a reply to its parent which often aggregates the replies from all the children and sends the aggregated result up the tree potentially involving aggregations on the way to the root which sends the overall response. The key point here is that each query needs to wait for the replies from either all the leaves (Web Search) or several hundreds of leaves (Key-value stores). Consequently, the overall response time of a query is affected by the slowest leaf so that the mean overall response time, and therefore the SLA budget, includes the $99^{th}$ - $99.9^{th}$ percentile leaf latency in a 1000-node cluster, known as the latency tail problem [11]. To maintain interactive user experience, the parents wait for replies only until the deadline and drop the replies that miss the deadline. Because the dropped replies affect response quality and revenue, OLDIs keep the fraction of missed deadlines low (e.g., 1%).

There is a wide variation in the leaves' reply latency due to variations in network and compute; as noted before, this variation is among the sub-queries within a query, not across queries. Requests from parents to leaves (and responses) may take varying time due to collisions at the packet buffers with the leaves' replies for multiple queries. Due to the tree-like software architecture and mostly balanced workload among the leaves, the leaves send their replies to the parent at about the same time; this phenomenon is called in-cast [5, 37, 38]. Because all the replies are destined for the same input port of the same node (parent), the replies are queued in the same packer buffer at the relevant datacenter network switch. Because in-casts are inevitable, the switches are provisioned with enough buffering to handle a few in-casts. However, the buffers are kept shallow for cost and latency reasons [5]. Therefore, multiple queries' in-casts occurring at about the same time and colliding at the buffers result in delays and buffer overflows; multiple queries are processed in parallel for high throughput. Further, there are also background flows from other applications on the cluster due to consolidation or to updating the OLDI data (e.g., Web index). Such collisions cause TCP time-outs and re-transmits resulting in the replies falling in the tail or exceeding the time budget. While such collisions are uncommon in general, they are common enough to affect the $90^{th}$-$99.9^{th}$ percentile latencies (e.g., in *every* query, 80% of replies incur 5 ms latency whereas the last 1% incur 20 ms). Further, such collisions are highly timing-dependent and therefore are highly unpredictable; the TCP-flow propagation delay for a leaf to realize that a collision has occurred is too long for the leaf to delay or slow down its sending rates (hence reactive schemes are unlikely to work).

While in-casts occur for replies, requests are also affected by a multiplexing strategy used to distribute the network load among most, if not all, of the datacenter's nodes. If the roles of the nodes serving as a parent or a leaf were fixed and unchanging, then the reply in-casts would cause hot spots in the network where the parent nodes would become repeated bottlenecks. To alleviate this problem, the role of a sub-tree parent for a query is randomized among the sub-tree's nodes i.e., a node is a parent for one query and a leaf for another. Such randomization ensures that in-casts are uniformly distributed among all the nodes [5]. We found in our simulations that using just one or two dedicated roots for 32 children exacerbates the reply in-casts and results in elongating the $99^{th}$ percentile of replies from around 22 ms with the randomization to 170 ms with 1-2 dedicated roots. Adding 4-8 dedicated roots performs as well as randomization but at 10-25% extra cost (i.e., 3-7 extra parents per 32 leaves). While randomization alleviates reply in-casts without extra cost, reply in-casts do occur, unfortunately, despite such randomization. Further, because the same node may issue a request as a parent to another node for one query and may send a reply as a leaf to the other node for another query, requests and replies can collide at the packet buffers. Consequently, requests caught in *unrelated* reply in-casts face delays and time-outs (the fractions are similar to those of replies as mentioned above).

Like the network, the compute in each leaf also exhibits latency variation due to work imbalance across queries despite good load balancing and hashing [34]. For instance, a Web Search query may lead to no matches at a leaf while finding many matches at another. Further, changes in the datacenter load also cause latency variation in compute. As such, compute latencies also vary by a wide range (e.g., in every query, 80% of leaves take 30 ms for compute including compute-queuing at the leaf server, whereas the last 1% take 70 ms).

Both in-casts and work imbalance occur at all loads. Higher loads increase the latency spread because queuing non-linearly dilates these latencies. In the case of compute-queuing delays, there are two effects: (1) queuing changes due to load changes, and (2) "instantaneous" changes in the work and load even at a fixed load.

## 2.2 Opportunities

In the presence of such variations, the average overall response time, and therefore the SLA budget, includes the tail latencies for the request-compute-reply-aggregate sequence. To account for compute-queuing delays, the tail latencies are measured at the expected peak load in a fully-provisioned datacenter. However, more than 80% of leaves complete well ahead of the deadline for every query (e.g., with 3-4x slack). TimeTrader targets this opportunity, the per-leaf per-request network slack and compute-queuing slack, which exists at all datacenter loads.

As discussed in Section 1, Pegasus [23] achieves load-proportional energy by slowing down leaf computation at lower loads based directly on response times. The paper shows that using response times is better than employing CPU-utilization-based dynamic voltage and frequency scaling (DVFS) which results in many missed deadlines because requests in the tail remain critical even at low loads. Pegasus uses datacenter-wide average response times as a measure of the load and uniformly slows down all the nodes at lower loads, while ensuring that SLA violations do not increase. Thus, Pegasus exploits the slack in the time budget, which is provisioned for the peak load, to *shift* the leaves' response time distributions (see Figure 2). In contrast, TimeTrader determines the slack for each individual leaf to reshape the response time distributions at all loads to be closer to the deadline than the default distribution (see Figure 2).

The Pegasus paper briefly describes a distributed version which uses individual server loading to determine the slowdown factors. It may seem that TimeTrader's compute-queuing slack arising from variations in instantaneous compute-queuing would be captured by this version (load-related slack in average queuing is already captured by the centralized version). While the paper suggests identifying high-load "hot" and low-load "cold" servers to modulate the factors, low average server loading over even fine time granularities does not ensure that most or all of the requests handled by a cold server have slack (i.e., individual leaf latencies are unpredictable). It is not clear that the requests with low slack would not miss their deadlines. Further, such load imbalance would be alleviated by careful re-distribution of the search index among the leaves, making persistent load imbalance over several queries unlikely even for short durations. Imbalance due to a few queries repeated numerously (i.e., popular search words) would be filtered by front-end caching of such popular queries to save cluster bandwidth.

The centralized version does not have this problem as it exploits the slack in datacenter-wide response times at lower loads as opposed that at higher loads without distinguishing among servers/leaves. Though this excellent paper has many insights and a detailed latency evaluation of the centralized version, the brief evaluation of the distributed version only compares estimated power savings using datacenter-wide load (centralized version) versus that using individual-server load (distributed version) but does not show latencies.

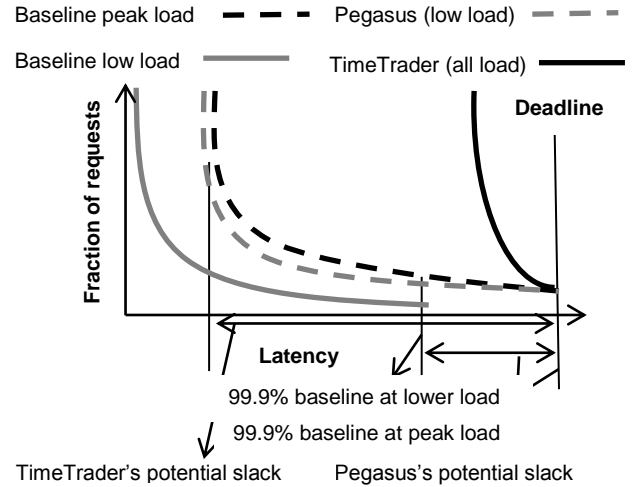

**Figure 2: Pegasus vs. TimeTrader**

## 2.3 Challenges

There are three issues in exploiting the sub-critical leaves' slack. First, though TimeTrader's opportunity exists at all loads, it is harder to exploit slack (i.e., to slow down) at higher loads. There may be slack in the requests as well as in instantaneous compute-queuing for TimeTrader even at higher loads, including the peak load. However, higher loads mean more queuing and TimeTrader's slack has to be distributed over the entire queue, and not just one request, to account for the fact that slower service affects all the queued requests and not just the one being slowed. In other words, any service slowdown is amplified by the queue length (e.g., $u^2/(1-u)$ in M/M/1 queues with a server utilization of $u$) so that the response time grows as the product of the slowdown factor and queuing. This interaction between queuing and service slowdown is the reason for TimeTrader's energy savings to decrease at higher loads. Nevertheless, TimeTrader still achieves significant energy savings even at the peak load. Note that the M/M/1 queue is just an example; datacenter nodes typically employ powerful multi-socket, multi-core servers and not uniprocessors.

Second, as discussed in Section 1, OLDIs have tight time budgets and are tail latency-limited. Because load variations at high loads cause compute-queuing and tail latencies to increase non-linearly, OLDIs usually operate well within the region where compute-queuing delays are kept low via throughput-parallelism. This condition implies that datacenters are provisioned well enough that even at the peak load there is compute-throughput slack. A key point here is that even though server utilizations are high at the peak load, high throughput parallelism ensures that the queuing delays are low (e.g., at 90% utilization, an *M/M/1* queue's response time is 10 * average service time whereas an *M/M/100* queue's response time is only 1.02 * average service time [18]). TimeTrader exploits this throughput slack to slow down sub-critical leaves without growth in the compute-queuing delays. Thus, TimeTrader maintains the same throughput as the baseline datacenter.

Finally, there is a subtle issue with OLDI time budget. For the SLA budget, the tail of the overall response latency matters and not the individual tail latencies of request, compute, or reply. In practice, to allow for independent development and optimizations of the network and compute parts, the total budget is broken into components for the network (request+reply) and compute. However, the chance of both a request and its reply hitting the tail is quite low and does not influence the $99^{th}$ percentile of the overall response latency. Consequently, the network's budget would account for the tail latency of the sum of the request and reply, and not the sum of the tail latency of each (i.e., the budget expects the risk of hitting the tail to be shared between the request and reply and essentially allows for the tail to be counted only once). This point implies that the request does not have a separate budget and therefore, the request slack cannot be known.

To address this issue, we choose to use separate budgets for request and reply. However, because of the risk sharing between request and reply, such separate budgets imply tighter individual budgets for the same total budget as the single-budget default. Indeed, our calculations show that considering two identical exponentially-distributed random variables, $X$ and $Y$, each of whose $99^{th}$ percentile is $v$, the $99^{th}$ percentile of $X+Y$ is $1.5v$ (single-budget case) whereas the $99^{th}$ percentile of $X$ + $99^{th}$ percentile of $Y$ is $2v$ (separate-budget case). Thus, for the same total budget, the separate budgets would each have to use $0.75v$ as the deadline to be met by the $99^{th}$ percentile.

Fortunately, this handicap is overcome by network optimizations specific to OLDIs which require separate budgets [37, 38]. These optimizations prioritize network flows for network bandwidth use based on each flow's deadline. The single-budget default cannot easily use these optimizations because (1) requests do not have a deadline and (2) request and reply are separate flows whose common budget would have to be communicated from the request to the reply via the compute layer while accounting for the lack of fine-grained clock synchronization between the nodes where the request and reply originate. We found that the separate-budget case employing the most recent of these optimizations, $D^2TCP$ [37], under the tighter, separate deadlines of $T/2$ achieves *fewer* missed deadlines than the single budget case under the single deadline of $T$. In the remainder of this paper, we use separate budgets for requests and replies, and employ $D^2TCP$ for all the systems we compare – baseline, Pegasus and TimeTrader.

## 2.4 Discussion

TimeTrader slows down the sub-critical leaves to save energy. While the leaf computation remains the same with or without TimeTrader (i.e., work is conserved), energy savings stems from the fact that executing at full speed and then idling till the next request is less efficient than executing at slower speed and idling less. Slower speeds save energy due to scaling of voltage (to whatever extent) and frequency. Idling consumes significant energy in fully-active mode; energy is lower in lower-power or sleep modes but OLDIs cannot exploit such modes because the sleep-to-active transitions are too long for OLDIs' time budgets and inter-arrival times [24, 25].

Finally, the slack uncovered by this paper can be used to save energy by slowing down leaf computation or to improve the quality of responses by increasing the computation. We explore the former option in this paper and leave the other options for future work.

## 3 TimeTrader

Recall from Section 1 that TimeTrader exploits the network slack in requests and individual queries' compute-queuing slack. TimeTrader slows down the individual, sub-critical leaves, to save energy without increasing SLA violations. To ensure that slowing down sub-critical requests does not hurt the critical requests that are queued behind the sub-critical requests, TimeTrader employs Earliest Deadline First (EDF) scheduling [22] that prioritizes the critical requests ahead of the sub-critical requests.

### 3.1 Request slack

Requests that arrive before their budgeted deadlines have slack which TimeTrader transfers to compute. Fortunately, because request comes before compute, this slack can be identified without prediction or the risk of missing the deadlines (recall from Section 2.1 that predicting network latencies is hard). However, requests originate at the parent node and compute occurs at a leaf, making it hard to accurately estimate the slack. Unfortunately, clock skew of several milliseconds between the parent and the leaf nearly rules out estimating slacks of similar magnitudes. Inter-node synchronization at such fine time granularity is hard [26, 28].

Instead of attempting to precisely determine the request slack, we use signals from the network about the presence or absence of packet drop and of imminent network congestion (typically due to an in-cast collision, as described in Section 2.1). Presence of these signals could mean no slack due to delays in the network whereas absence confirms some slack. While there may still be some slack even in the former case, we conservatively assume there is none. Because congestion is uncommon in datacenters that host OLDIs, our conservative assumption does not degrade our savings.

Determining the exact slack amount in the absence of the signals involves two cases: packet drop and imminent congestion. The former case results in retransmission which is marked by the sender (parent) with a packet header bit. The receiver (leaf) then assumes no slack. In the absence of retransmission, there is slack of one minimum timeout duration (TCP's $RTO_{min}$) based on the facts that any retransmission occurs only after a timeout and that network tail latency typically includes $RTO_{min}$ to cover one timeout due to in-cast collisions (Section 2.1). Consequently, we conservatively set the request slack to be $RTO_{min}$; there is natural padding of around 5 ms in the budgets to account for protocol overheads (e.g., $RTO_{min}$ of 20 ms is commonly used on datacenters [5]). The latter case of imminent congestion is signaled by Explicit Congestion Notification (ECN) [32]. Network switches detect imminent congestion when packet buffers are occupied above certain watermarks signifying queuing delays, and use ECN bits in packet headers to pass this information. Upon receipt, the leaf assumes no slack. In the absence of ECN markings, we determine the slack amount by empirical evaluation of network delays in the presence of ECN markings. In our experiments, we set this slack to be *request budget – median network latency*.

### 3.2 Individual compute-queuing slack

Compute-queuing slack stems from variations in the queuing at the leaf. Like requests, queuing comes before the actual compute and therefore queuing slack can be identified without prediction or the risk of missing the deadlines. Pegasus exploits the datacenter-wide average queuing slack (i.e., budget – average queuing), which is present at lower loads (the compute budget is determined by the queuing delay at the peak load). In contrast, we exploit individual request's queuing slack based on the fact that even under a fixed load, queuing varies from one request to another.

To determine this slack, we determine the queuing time by timestamping the arrival of a request and the start of computation at the leaf (both arrival and computation occur at the same server so there are no clock skew issues). The compute-queuing slack is the average queuing delay at the peak load minus the given request's actual queuing delay. The former is pre-determined empirically; and the latter depends on the current load and variations in queuing seen by the current request and is measured via the timestamping. Thus,

*compute-queuing slack = average peak wait – current wait*

*total slack = request slack + compute-queuing slack*

As discussed in Section 2.3, this total slack has to be attenuated (i.e., scaled) before being applied as a slowdown to account for the fact that slower computation affects all the queued requests and not just the current request. One other subtle issue is that going to a lower power setting in CPUs requires choosing a slowdown factor. While we know the total slack amount, we do not know how long the current request will take and therefore, we cannot compute a slowdown factor. Fortunately, both these issues – attenuation and unknown service time – can be addressed by observing that the compute budget accounts for worst-case queuing delays and worst-case service times. Further, some slack is spent in RAPL latency. Therefore, we set

*slowdown =(total slack – $RAPL_{latency}$)\*scale/compute budget*

where *scale* is a factor to further moderate the slowdown. Scale depends on both load and applications (i.e., service time distributions and budgets). Higher load implies lower value for *scale* to reduce the slowdown factor and impact on throughput. Instead of using statically configured *scale* values for each application, we employ a simple control algorithm that dynamically determines *scale* by monitoring the percentage of missed deadlines at each leaf server every 5 seconds. If the percentage of missed deadlines in the current interval is less than the SLA target by more than 5% (i.e., there is 5% room in the budget), we increase *scale* by 0.05. Else, we reduce *scale* by 0.05 until there is room or the *scale* is 0. Thus, there is a guard band of 5% to avoid SLA violations. Even at the peak load, there is room to exploit. However, Pegasus cannot exploit this room because it does not distinguish critical requests from sub-critical requests, at the *same* leaf server. TimeTrader saves energy even at the peak load by slowing down sub-critical requests using a non-zero *scale* value without directly affecting critical requests that have 0 *total slack* (*scale* does not matter). *Further,* EDF shields critical requests from the queuing effects that arise from the slowing down of sub-critical requests. *Thus,* by using per-request slack and EDF, TimeTrader saves energy at all loads. Table 1 shows *scale* values across various loads for *Search* and *memcached*.

**Table 1: Values for *scale***

| Utilization | WebSearch | Memcached |
|---|---|---|
| 30% | 0.7 | 0.8 |
| 60% | 0.4 | 0.5 |
| 90% | 0.2 | 0.2 |

To set the core's speed as per the slowdown factor, we employ RAPL [1], which requires less than 1 millisecond, making it suitable for OLDI timescales. One issue is that modern processors are multicores with hardware multithreading (i.e., Simultaneous Multithreading (SMT) [36]). Multiple cores may be processing either multiple requests of the same query or different queries, and in either case the slack for the cores may be different. Further, each core may have a few SMT contexts for each of which the slack may be different. To address this issue, we assume that each core's power settings can be controlled independently of other cores' settings. While current offerings of RAPL control only the overall package power, individual core control is a relatively small extension and is likely to be implemented in the near future. To address the SMT contexts within a core, we conservatively use the worst of the contexts' individual slowdown factors to avoid violating deadlines. Because the number of SMT contexts per core is only a few (e.g., 2), this conservative assumption – i.e., the worse of two slowdown factors – does not diminish our opportunity.

When we explored slowing down main memory in addition to the CPU, the fact that memory is shared among all the cores of a server severely limits the memory slowdown factor in the presence of such a conservative assumption. For instance, for a 32-core server, the memory slowdown factor would have to be the worst among all the 32 cores' factors, which would likely be zero. Therefore, we slow down only the cores and not memory. Nonetheless, because CPUs contributes about 60% of server power [8], our opportunity remains significant.

### 3.3 Deadline-based compute-queuing

Recall from Section 1 that the presence of slack is not sufficient to guarantee avoiding missing of the deadlines. Slowing down a sub-critical request which has slack may hurt another critical request that is queued behind the sub-critical request. To address this issue, we exploit Earliest Deadline First (EDF) scheduling that decouples critical requests from the queuing delays of sub-critical requests by placing the former ahead of the latter in the leaf server's queues.

The decoupling is not perfect due to the fact that arriving critical requests may still see elongated, residual service times of sub-critical requests in the absence of pre-emption (whose delays would not be suitable in our context of tight deadlines). Nevertheless, the decoupling enables EDF to pull in the tail and to reshape the leaves' response time distribution; the mean response time does not improve because as critical requests' response times get shorter the sub-critical requests' times get longer. However, EDF enables TimeTrader to use per-leaf slack to slow down sub-critical requests, thereby *further* shifting the distribution closer to the deadline. Though such slow down lengthens the mean service time, such an increase taps into the throughput slack described in Section 2.3 and hence does not worsen throughput. Still, the throughput slack may not be enough to exploit the full total slack in which case we give up some energy savings to avoid throughput loss.

OLDI implementations typically use well-defined APIs which cleanly separate request queue management and thread computation modules (e.g., work-stealing task queues). EDF is typically available with standard queue management libraries (e.g., pthread_set_schedparam() can be used to achieve EDF by setting the priority to be the deadline) and adds negligible overhead (section **5**). As such, the libraries enable TimeTrader to be used easily in a host of OLDIs.

## 4 Methodology

TimeTrader involves three aspects: network latency, compute latency, and compute power. We use real-system measurements for compute latency and compute power, a rack-scale real implementation to show proof-of-concept, and at-scale simulations for network latency. The compute aspects involve only one server because over long periods of time all servers are statistically identical in response times and power consumption and hence real-system measurements are feasible. Further, because tail effects are more pronounced in large clusters (e.g., 1000 node) to which we do not have access, we rely on simulations to study the network aspect.

**Benchmarks:** We simulate two OLDI benchmarks, *Web Search (Search)* and *memcached* (key-value store), from CloudSuite 2.0 [13]. We modify the *memcached* driver to look up a batch of objects in each request, with an average batch size of 50 as is typical [27], instead of single objects as done in CloudSuite. We generate *Search*'s index from Wikipedia and *memcached*'s objects from Twitter. In our runs, *Search* and *memcached*, respectively, support peak queries-per-second rates of 3000 and 20,000 using 100 threads per leaf server at 90% utilization (corresponding to a modern server with 4 sockets, 12 cores per-socket, and 2 SMT contexts per core). Our *memcached* throughput of 20,000 queries-per-second with a batch size of 50 objects (i.e., 1 M objects/s) matches the throughputs reported in [27]. These threads provide high throughput parallelism to match the peak load (i.e., the threads are copies processing the same index/key-value slice and not separate leaves processing different slices).

The benchmarks use a parent-to-leaf fan-out of 32 (a standard value). For each query, we randomly choose a node to be the parent (Section 2.1). We set the budgets as: total 200 ms, request 25 ms, reply 25 ms, leaf compute 75 ms (*Web Search*) and 20 ms (*memcached*), and aggregate and remaining network (aggregate-root communication) 75 ms. The network and compute budgets are the $99^{th}$ percentile latencies achieved by, respectively, our network using $D^2TCP$ and compute nodes at the peak load. We target less than 1% missed deadlines (i.e., these deadlines are tight and do not offer any "easy" opportunity for TimeTrader). The network and compute budgets are in line with [5, 37, 38] and [34], respectively. TimeTrader focuses on request, compute and reply for a total of 125 ms (*Web Search*) and 70 ms (*memcached*) which are the deadlines in our experiments. We use request sizes of 2 KB and reply sizes of 16-64 KB chosen uniformly randomly, and background flow sizes of 1 and 10 MB chosen uniformly randomly (Section 2.1); the total traffic is split evenly between OLDI and background flows. These message characteristics match publicly-available distributions from production OLDIs [9]. In all our experiments, the network utilization is 20% which is realistic for datacenters [5] (i.e., the network is over-provisioned and yet incurs in-cast collisions).

**Real Implementation**: Our real implementation uses 9 servers (8 leaves and 1 parent, with a fan-out of 8), which are connected to a rack switch using 1 Gbps links. We implement TimeTrader's slack computations and EDF at the leaf servers for *Search*. We distribute the index to all the leaf servers. We vary the query rate using Faban (CloudSuite). Because our switches do not support ECN, we timestamp requests at the parent and leaf servers to infer request slack because clock drifts are not a problem at this scale (i.e., the clocks drift by at most 200 microseconds during our evaluation). We generate background traffic between servers (i.e., all-to-all traffic) using Iperf [2] to maintain a network utilization of 20% (i.e., 200 Mbps). This traffic provides incast effect at rack scale. Finally, we reduce the request budget from 25 ms to 15 ms because tail effects (i.e., incast) are less intense at small scale. Therefore, our budgets are not over-provisioned.

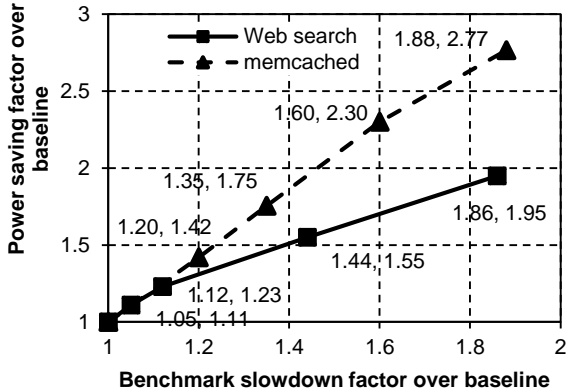

Figure 4: Power-Latency relationship

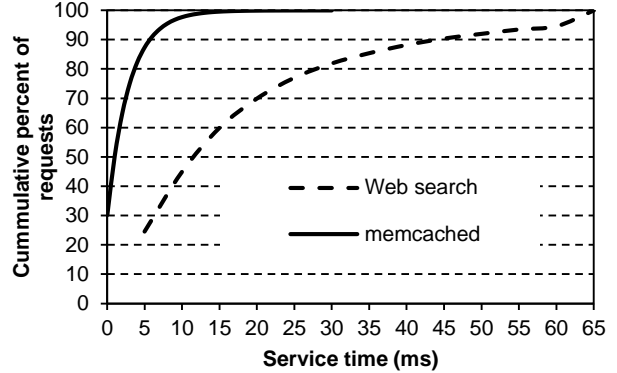

Figure 3: Service Time distributions

**Compute latency and power:** To measure compute latency and power, we run the benchmarks on a system using an Intel IvyBridge-based CPU. We generate a leaf compute latency distribution (service time only without any compute-queuing delay) for our benchmarks running on the system (see Figure 3). The compute latency distribution confirms the wide spread of compute latencies. The compute time for *search* is significant whereas that for *memcached* is shorter (object lookups are fast) making *memcached* network-limited and providing more opportunity for slowing down compute. The compute budgets for *search* and *memcached* at 75 ms and 20 ms are slightly more than the 99th percentile latencies to account for queuing delays at the peak load.

Using RAPL, we vary the CPU clock speed from 2.5 GHz to 1.2 GHz and obtain per-request latency (total latency, not just clock speed) and per-core power. Figure 4 shows active power saving factor (Y axis) and request slowdown factor (X axis) for *search* and *memcached*; active power = total power − idle power. As the slowdown increases, the power savings are slightly super-linear over compute slowdown in the beginning where there may be some voltage scaling and then the savings slightly flatten when voltage cannot scale as much. We use these compute latency and total power values (including idle) with network latency to report power and performance.

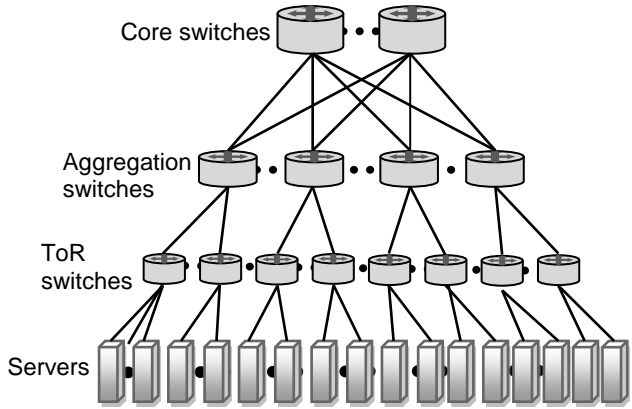

Figure 5: Simulated Network

**Network latency:** Using *ns-3* [3], a widely-used simulator, we simulate the network depicted in Figure 5, which uses a fat-tree topology typical of datacenter networks [4]. There are 64 racks with each rack having up to 16 servers (i.e., a 1000-server cluster). Each server connects to the top-of-rack (ToR) switch via a 10 Gbps link. Going up from the ToR level, there is a bandwidth over-subscription of 2x at each level, as is typical [4]. We sized the packet buffers in the ToR switches to match typical buffer sizes of shallow-buffered switches in real data centers (4MB) [5]. We set the link latencies to 20 μs, achieving an average of round-trip time (RTT) of 200 μs, which is representative of datacenter network RTTs. To reduce the effects of in-cast collisions, we add a 1-ms jitter to each leaf's reply [14].

To simulate a deadline-aware TCP implementation that exploits the separate request-reply budgets (Section 2.3), we use $D^2TCP$ [37] on top of ns-3's TCP New Reno protocol [2]. (code obtained from $D^2TCP$'s authors). All $D^2TCP$ parameters (e.g., deadline imminence factor) match those in [37] and are available with the code. We set $RTO_{min}$ for all the protocols to be 20 ms. We use the same separate request-reply budgets and $D^2TCP$ in all the systems we compare – baseline (no power management), Pegasus and TimeTrader. The latencies we observe closely match those reported in other papers, including production runs [37].

**All together:** In *ns-3*, we simulate TimeTrader's EDF scheduling (Section 3.3) and compute the total slack as a function of the request slack and compute-queuing slack (Section 3.2). We also simulate Pegasus to determine its slack based on the datacenter-wide load as compared to the peak. We apply TimeTrader's total slack and Pegasus's slack as slowdown factors to our real-system runs to measure TimeTrader's and Pegasus's energy savings.

## 5 Rack-scale implementation results

We validate TimeTrader's energy gains using a real rack-scale implementation and quantify its overheads.

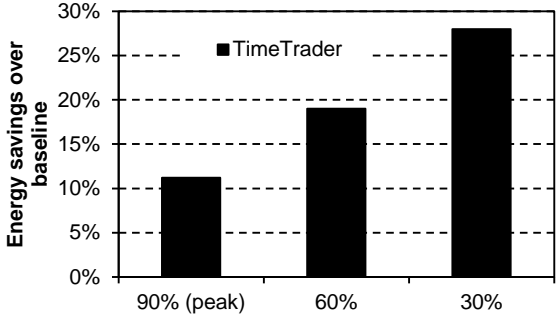

Figure 6: Rack-scale Energy Savings

Figure 6 shows our energy savings for *Search* over a baseline without power management. The Y axis shows energy savings (including idle) and the X axis shows *Search* running at 90% (peak), 60%, and 30% load. Our slowdowns of 7%, 16%, and 27% (not shown) correspond to energy savings of 11%, 19%, and 28% (shown in Figure 6) at 90%, 60% and 30% load. Because, tail effects are less intense at rack scale, our energy savings are less than our savings at-scale (section 0). Nevertheless, TimeTrader's energy savings are still significant.

Further, we use the real implementation to measure the overhead of EDF and timestamping (i.e., needed for determining compute-slack). We find that EDF adds an overhead of 330 microseconds for re-prioritizing about 15 entries (i.e., our $99^{th}$ percentile queue length). Timestamping (i.e., used for calculating compute-slack) adds an additional overhead of 45 microseconds per request. These overheads are negligible compared to OLDI service times, which are in the order of tens of milliseconds.

## 6 At-scale simulation results

Now we show our at-scale results. We start with comparing the energy savings of TimeTrader and Pegasus, the main result of the paper. We explain the savings by presenting the distributions of (a) request slack, (b) compute-queuing slack, and (c) the request-compute-reply latency. We then show a binning of requests based on their CPU core's power state TimeTrader and Pegasus. Finally, we isolate the contributions of EDF, request slack, and compute slack.

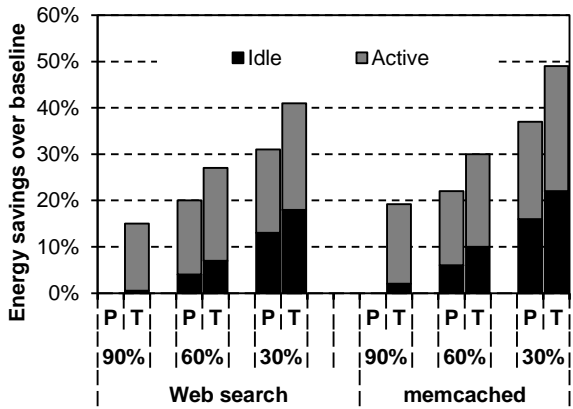

Figure 8: At-Scale Energy Savings

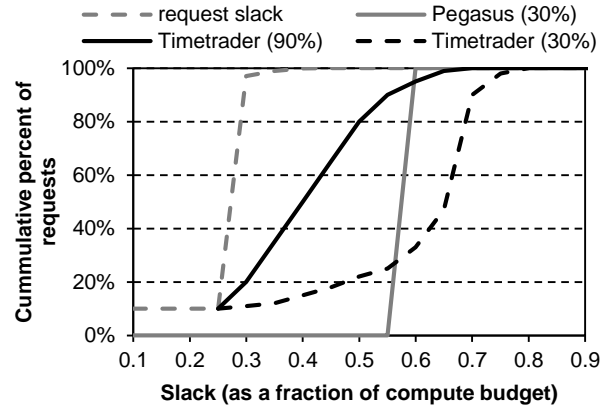

Figure 7: Slack distribution for *search*

### 6.1 Energy savings

Figure 8 compares the energy savings of Pegasus and TimeTrader over a baseline cluster without power management. The Y axis shows the total energy savings (including idle) and the X axis shows the benchmarks running at 90% (peak), 60%, and 30% load with "*P*" and "*T*" denoting Pegasus and TimeTrader, respectively. In all the three systems, less than 1% of queries exceed the 125-ms (*search*) and 70-ms (*memcached*) request-compute-reply budgets (i.e., they all meet our target of less than 1% missed deadlines). Because Pegasus does not save energy at the peak load, that bar is zero.

Both Pegasus and TimeTrader achieve significant savings at low loads with TimeTrader achieving more due to the difference between Pegasus's datacenter-wide average loads based slack versus TimeTrader's per-query, per-leaf slack. For instance, at 30% load, TimeTrader achieves around 42% (*search*) and 49% (*memcached*) savings compared to Pegasus's 32% and 37%; these savings amount to improvements of 17% (0.68/0.58) and 24% (0.63/0.51) over Pegasus. Both systems save more in *memcache* than in *search* because *memcache*'s shorter compute latency than network latency allows longer slacks and greater slowdown factors. By slowing down, Pegasus and TimeTrader save both active and idle energy (Section 2.4). As the load increases, idle power savings increase as expected due to less idling. Further, TimeTrader saves more than 15% energy at the peak load during which the power consumption is more than twice than that during 30% load (it is misleading to compare the savings percentages at different loads which correspond to different amounts of power consumption). Because datacenter loads are moderate to high during half the day (diurnal pattern), TimeTrader's savings are significantly higher than Pegasus's.

### 6.2 Slack and latency distributions

To explain these savings, we plot the slack for *search* in Pegasus and TimeTrader in Figure 7. We do not show

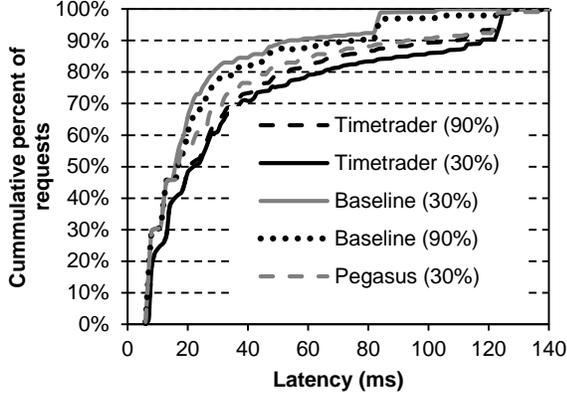

Figure 10: Request-Compute-Reply latency for Search

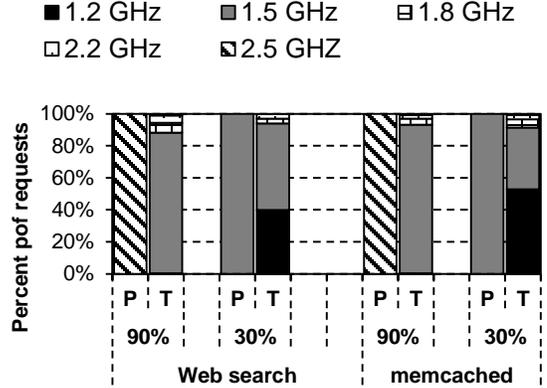

Figure 9: Power states distribution

*memcached's* slack which is similar. The X axis shows the slack as a fraction of the compute budget and the Y axis shows the cumulative percent of requests. We show the request slack (relevant only for TimeTrader), TimeTrader's total slack at 90% and 30% loads, and Pegasus's total slack at 30% load (zero at 90% load, not shown). The request slack is the same at all loads because the network is over-provisioned (Section 4) [5]. We do not show 60% load to avoid cluttering the graph.

Almost the entire request slack is available to 90% of the requests in TimeTrader because in-casts are infrequent (Section 2.1). The difference between the request slack and TimeTrader's total slack is the compute slack (both loads). In TimeTrader, even at 90% load, 90% of requests have a slack of (0.25 * compute budget) or more, confirming that most requests are sub-critical even at the peak load; at 30% load, 80% of requests have a slack of (0.5 * compute budget) or more. Further, Pegasus's slack at 30% load corresponds to the difference in the 99th percentile latencies for 30% load and 90% load (peak), and is available to almost all requests (i.e., Pegasus's slack is mostly a function of the load and does not vary from one request to another for a fixed load). Compared to Pegasus, at 30% load, TimeTrader has lower slack for 10% of requests because TimeTrader exploits per-request slack where a higher slack for one request sometimes increases the queuing delay for another request cutting into the latter's slack (i.e., there is some give-and-take among the requests). These values are the total slack whereas TimeTrader's slowdown factors involve another scaling factor to moderate for the load (Section 3.2 and Table 1). Nevertheless, TimeTrader's longer slack results in higher energy savings.

The slowdown factors for Pegasus and TimeTrader closely follow the slack amounts in Figure 7. We note that by carefully exploiting the throughput slack, TimeTrader maintains the same throughput as the baseline at all loads (fall in throughput would manifest as many missed deadlines).

To illustrate that TimeTrader reshapes the request-compute-reply latency distribution while Pegasus shifts the distribution, we plot the latency distributions for *search* in Figure 10. The plot shows the distributions for the baseline, TimeTrader, and Pegasus at 30% and 90% load (Pegasus at 90% coincides with the baseline at 90%). We note that the plot shows the total latency including the reply component to show the overall effect of the schemes, as opposed to Figure 7 which shows only request and compute components. As expected, TimeTrader reshapes the distributions at both loads, albeit more at 30% than 90% due to greater latency and throughput slacks. In contrast, Pegasus shifts the baseline curve at 90% load to the right when the load is 30%. Also, as load increases, the systems diverge more at higher percentiles than at lower percentiles. Because OLDIs' M/M/96 queues, unlike M/M/1 queues, exhibit highly non-linear queuing – higher percentiles of queuing delay increase more abruptly than lower percentiles at higher loads.

### 6.3 Power states

To understand TimeTrader's energy savings, we bin the requests based on the CPU core's power state for each request. Each power state corresponds to a core clock speed which is scaled based on the request's slowdown factor. Figure 9 shows the fraction of requests in each bin for Pegasus (P) and TimeTrader (T) at 90% (peak) and 30% loads running *search* and *memcached*. The bins span 1.2 GHz to 2.5 GHz.

We consider *search* first. Pegasus does not slow down requests at 90% load and incurs the highest clock speed and power. In contrast, TimeTrader even at 90% load slows down 85% of the requests by 20% or more which corresponds to the second-slowest state (1.5 GHz) (Figure 9). As the load decreases to 30% and the slack increases, Pegasus also slows down requests to the same state. However, TimeTrader uses the slowest state for many requests (40%) and saves more energy. In contrast to TimeTrader's per-query metrics, Pegasus's datacenter-wide average metrics imply that for a fixed load the power states do not change much. The trends in memcache are similar.

### 6.4 Isolation of impact

We isolate the impact of EDF, request slack, and compute slack on TimeTrader's energy savings. Figure 11 shows the four systems' energy savings over the baseline: TimeTrader without EDF, TimeTrader using only request slack and EDF, TimeTrader using only compute slack and EDF, and TimeTrader (whole). As before, all the systems have the same time budget and target of missed deadlines (1%). The X axis shows 90% and 30% load and our benchmarks.

Without EDF, critical requests queued behind slowed-down sub-critical requests are likely to be affected. To achieve the same percent of missed deadlines, TimeTrader's slowdown factors are greatly reduced. Hence, without EDF, TimeTrader's savings are modest though they grow as the load decreases from 90% to 30% due to the availability of more slack. TimeTrader using only request slack achieves a significant fraction of that of TimeTrader (whole) at 90% load where compute slack is limited and this fraction diminishes as the load decreases to 30%. As expected, this trend reverses for TimeTrader using only compute slack.

## 7 Related work

Previous work on improving energy efficiency fall into the following four categories: datacenter power management, software consolidation, exploiting low-power modes, and real-time systems.

In the first category, a datacenter-wide power budgeting approach [33] allows the budget to be shared among multiple entities (e.g., racks and servers) to achieve high power-supply utilization and efficiency, analogous to chip-level power budget management in [17]. A coordinated power management approach [30] integrates several power controllers to avoid conflicting decisions and improve overall efficiency.

The second category of software consolidation improves energy efficiency by consolidating workload on under-utilized servers so that the servers operate at high utilization levels which are also energy efficient. While consolidation of batch workloads such as MapReduce [10, 19] and multi-programmed workloads [12] is possible, OLDIs' tight latency budgets and large memory footprints disallow such consolidation. Bubble-flux [39] shows that OLDIs can be co-located with batch jobs under looser latency budgets but improving the utilization is hard under tighter budgets.

Exploiting low-power modes, the third category, proposes low-power idle states or leverages turning servers off (e.g., PowerNap [24], Blink [35]). However, the transition times are too long for the tight OLDI latency budgets; and OLDIs need all the leaf servers to stay turned on. Other work [25] studies OLDI workloads and concludes that the tight budgets necessitate a cluster-wide approach to power management, similar to Pegasus and TimeTrader. We have extensively discussed and contrasted the two schemes. Other proposals employ DVFS to improve throughput-centric batch workloads [17, 20, 31]. However, these proposals do not address OLDI's latency constraints.

In the fourth category, real-time systems have tight latency constraints like OLDIs so that energy efficiency can be achieved via DVFS by slowing down based on the jobs' deadlines [6, 21, 29]. However, these proposals exploit real-time jobs' characteristics that are significantly different from those of OLDIs (e.g., apriori knowledge of number and duration of jobs running single-node systems). OLDIs do not permit such apriori knowledge and are distributed applications running on large clusters.

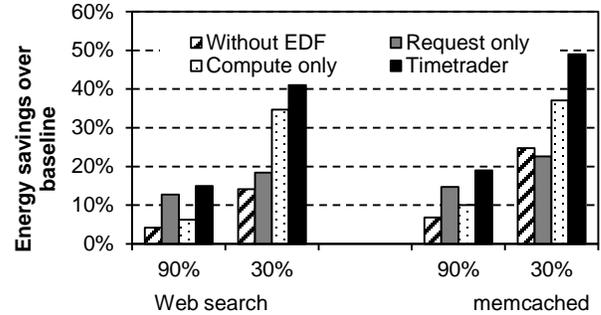

**Figure 11: Impact of EDF, request slack, and compute slack**

Finally, we have discussed many networking proposals targeting the in-cast problem in OLDIs [5, 37, 38]. These proposals address only network latency and do not explore dynamically sharing the latency budget between network and compute, as done by TimeTrader.

## 8 Conclusion

Reducing the energy of datacenters running on-line, data-intensive applications (OLDIs) is challenging due to OLDIs' tight response time requirements. In OLDIs, each user query goes to all or many of the nodes in the cluster, so that overall time budget is dictated by the tail of the replies' latency distribution; replies see latency variations both in the network and compute. We proposed *TimeTrader* to reduce energy by exploiting sub-critical replies' latency slack. While previous work *shifts* the leaves' response time distribution to consume the slack at lower loads, TimeTrader *reshapes* the distribution at all loads by slowing down individual sub-critical nodes without increasing missed deadlines. TimeTrader exploits slack in both the network and compute budgets. Further, TimeTrader leverages Earliest Deadline First scheduling to decouple critical requests from the queuing delays of sub-critical requests which can then be slowed down without hurting critical requests. Using a combination of real-system measurements and at-scale simulations, we showed that without adding to missed deadlines, TimeTrader saves 15-49% energy in a datacenter with 512 nodes, whereas previous work saves 0% and 31-37%.

By exploiting latency slack in the highly-latency-sensitive OLDIs, TimeTrader converts OLDIs' performance disadvantage of latency tails into an energy advantage. As OLDIs grow in scale due to the ever-increasing data and in importance due to the ever-growing number of OLDI-reliant services, energy consumption will become only more important. As such, techniques like TimeTrader will be important in the march towards energy efficiency.

# References

1. Intel® 64 and IA-32 Architectures Software Developer Manuals *Systems Programming Guide, part 2*, 2013.
2. Iperf - The TCP/UDP Bandwidth Measurement Tool *https://iperf.fr/*.
3. The ns-3 discrete-event network simulator, http://www.nsnam.org/.
4. Al-Fares, M., Loukissas, A. and Vahdat, A. A scalable, commodity data center network architecture *Proceedings of the ACM SIGCOMM 2008 conference on Data communication*, ACM, Seattle, WA, USA, 2008.
5. Alizadeh, M., Greenberg, A., Maltz, D.A., Padhye, J., Patel, P., Prabhakar, B., Sengupta, S. and Sridharan, M. Data center TCP (DCTCP) *Proceedings of the ACM SIGCOMM 2010 conference*, ACM, New Delhi, India, 2010.
6. Aydin, H., Melhem, R., Moss, D., Mej, P. and a, A. Power-Aware Scheduling for Periodic Real-Time Tasks. *IEEE Trans. Comput.*, *53* (5). 584-600.
7. Barroso, L.A., Dean, J. and Holzle, U. Web Search for a Planet: The Google Cluster Architecture. *IEEE Micro*, *23* (2). 22-28.
8. Barroso, L.A. and Hölzle, U. *The Datacenter as a Computer: An Introduction to the Design of Warehouse-Scale Machines*. Morgan and Claypool, 2009.
9. Benson, T., Akella, A. and Maltz, D.A. Network traffic characteristics of data centers in the wild *Proceedings of the 10th ACM SIGCOMM conference on Internet measurement*, ACM, Melbourne, Australia, 2010.
10. Chen, Y., Alspaugh, S., Borthakur, D. and Katz, R. Energy efficiency for large-scale MapReduce workloads with significant interactive analysis *Proceedings of the 7th ACM european conference on Computer Systems*, ACM, Bern, Switzerland, 2012.
11. Dean, J. and Barroso, L.A. The tail at scale. *Commun. ACM*, *56* (2). 74-80.
12. Delimitrou, C. and Kozyrakis, C. Paragon: QoS-aware scheduling for heterogeneous datacenters *Proceedings of the eighteenth international conference on Architectural support for programming languages and operating systems*, ACM, Houston, Texas, USA, 2013.
13. Ferdman, M., Adileh, A., Kocberber, O., Volos, S., Alisafaee, M., Jevdjic, D., Kaynak, C., Popescu, A.D., Ailamaki, A. and Falsafi, B. Clearing the clouds: a study of emerging scale-out workloads on modern hardware *Proceedings of the seventeenth international conference on Architectural Support for Programming Languages and Operating Systems*, ACM, London, England, UK, 2012.
14. Floyd, S. and Jacobson, V. The synchronization of periodic routing messages *Conference proceedings on Communications architectures, protocols and applications*, ACM, San Francisco, California, USA, 1993.
15. Google. Efficiency: How we do it *http://www.google.com/about/datacenters/efficiency/internal/*.
16. Hoff, T. Latency is Everywhere and it Costs You Sales - How to Crush it *http://highscalability.com/blog/2009/7/25/latency-iseverywhere-and-it-costs-you-sales-how-to-crush-it.html.*, 2009.
17. Isci, C., Buyuktosunoglu, A., Cher, C.-Y., Bose, P. and Martonosi, M. An Analysis of Efficient Multi-Core Global Power Management Policies: Maximizing Performance for a Given Power Budget *Proceedings of the 39th Annual IEEE/ACM International Symposium on Microarchitecture*, IEEE Computer Society, 2006.
18. Kleinrock, L. *Theory, Volume 1, Queueing Systems*. Wiley-Interscience, 1975.
19. Lang, W. and Patel, J.M. Energy management for MapReduce clusters. *Proc. VLDB Endow.*, *3* (1-2). 129-139.
20. Lee, J. and Kim, N.S. Optimizing throughput of power- and thermal-constrained multicore processors using DVFS and per-core power-gating *Proceedings of the 46th Annual Design Automation Conference*, ACM, San Francisco, California, 2009.
21. Lin, C. and Brandt, S.A. Improving Soft Real-Time Performance through Better Slack Reclaiming *Proceedings of the 26th IEEE International Real-Time Systems Symposium*, IEEE Computer Society, 2005.
22. Liu, C.L. and Layland, J.W. Scheduling Algorithms for Multiprogramming in a Hard-Real-Time Environment. *J. ACM*, *20* (1). 46-61.
23. Lo, D., Cheng, L., Govindaraju , R., Barroso, L.A. and Kozyrakis, C. Towards Energy Proportionality for Large-Scale Latency-Critical Workloads *The 41st Annual International Symposium on Computer Architecture*, Minnesota, MN, 2014, 301-312.
24. Meisner, D., Gold, B.T. and Wenisch, T.F. PowerNap: eliminating server idle power *Proceedings of the 14th international conference on Architectural support for programming languages and operating systems*, ACM, Washington, DC, USA, 2009.
25. Meisner, D., Sadler, C.M., Andr, L., Barroso, Weber, W.-D. and Wenisch, T.F. Power management of online data-intensive services *Proceedings of the 38th annual international symposium on Computer architecture*, ACM, San Jose, California, USA, 2011.
26. Moon, S.B., Skelly, P. and Towsley, D., Estimation and removal of clock skew from network delay measurements. in *INFOCOM '99. Eighteenth Annual Joint Conference of the IEEE Computer and Communications Societies. Proceedings. IEEE*, (1999), 227-234.
27. Nishtala, R., Fugal, H., Grimm, S., Kwiatkowski, M., Lee, H., Li, H.C., McElroy, R., Paleczny, M., Peek, D., Saab, P., Stafford, D., Tung, T. and Venkataramani, V. Scaling Memcache at Facebook *Proceedings of the 10th USENIX conference on Networked Systems Design and Implementation*, USENIX Association, Lombard, IL, 2013.
28. Paxson, V. On calibrating measurements of packet transit times *Proceedings of the 1998 ACM SIGMETRICS joint international conference on Measurement and modeling of computer systems*, ACM, Madison, Wisconsin, USA, 1998.
29. Pillai, P. and Shin, K.G. Real-time dynamic voltage scaling for low-power embedded operating systems *Proceedings of the eighteenth ACM symposium on Operating systems principles*, ACM, Banff, Alberta, Canada, 2001.
30. Raghavendra, R., Ranganathan, P., Talwar, V., Wang, Z. and Zhu, X. No "power" struggles: coordinated multi-level power management for the data center *Proceedings of the 13th international conference on Architectural support for programming languages and operating systems*, ACM, Seattle, WA, USA, 2008.
31. Rajamani, K., Rawson, F., Ware, M., Hanson, H., Carter, J., Rosedahl, T., Geissler, A., Silva, G. and Hua, H. Power-performance management on an IBM POWER7 server *Proceedings of the 16th ACM/IEEE international symposium on Low power electronics and design*, ACM, Austin, Texas, USA, 2010.
32. Ramakrishnan, K., Floyd, S. and Black, D. *The Addition of Explicit Congestion Notification (ECN) to IP*. RFC Editor, 2001.
33. Ranganathan, P., Leech, P., Irwin, D. and Chase, J. Ensemble-level Power Management for Dense Blade Servers *Proceedings of the 33rd annual international symposium on Computer Architecture*, IEEE Computer Society, 2006.
34. Ren, S., He, Y. and McKinley, K. A Theoretical Foundation for Scheduling and Designing Heterogeneous Processors for Interactive Applications *the 11th International Conference on Autonomic Computing (ICAC 14)*, USENIX Association, Philadelphia, PA, 2014.
35. Sharma, N., Barker, S., Irwin, D. and Shenoy, P. Blink: managing server clusters on intermittent power *Proceedings of the sixteenth international conference on Architectural support for programming languages and operating systems*, ACM, Newport Beach, California, USA, 2011.

36. Tullsen, D.M., Eggers, S.J. and Levy, H.M. Simultaneous multithreading: maximizing on-chip parallelism *Proceedings of the 22nd annual international symposium on Computer architecture*, ACM, S. Margherita Ligure, Italy, 1995.
37. Vamanan, B., Hasan, J. and Vijaykumar, T.N. Deadline-aware datacenter tcp (D2TCP) *Proceedings of the ACM SIGCOMM 2012 conference on Applications, technologies, architectures, and protocols for computer communication*, ACM, Helsinki, Finland, 2012.
38. Wilson, C., Ballani, H., Karagiannis, T. and Rowtron, A. Better never than late: meeting deadlines in datacenter networks *Proceedings of the ACM SIGCOMM 2011 conference*, ACM, Toronto, Ontario, Canada, 2011.
39. Yang, H., Breslow, A., Mars, J. and Tang, L. Bubble-flux: precise online QoS management for increased utilization in warehouse scale computers *Proceedings of the 40th Annual International Symposium on Computer Architecture*, ACM, Tel-Aviv, Israel, 2013.